# Spatial asymmetry of photoelectron emission in strong-field ionization of atoms irradiated by a bichromatic laser field

**P. E. Pyak♣, E. A. Andreeva and V. I. Usachenko**

Physics Department, National University of Uzbekistan, Tashkent 100174, Uzbekistan

**Abstract** The phenomenon of spatial (polar) asymmetry of produced photoelectron emission due to strong-field multiphoton process of above-threshold ionization (ATI) in atoms irradiated by a linearly polarized two-color (bichromatic) laser field consisting of coherent superposition of two commensurate harmonics (of frequencies $\omega_1$ and $\omega_2$ is considered and studied theoretically. The problem is addressed within the conventional *strong-field approximation* (SFA) in standard *velocity gauge* (VG) formulation under condition of arbitrary value of the so-called Keldysh parameter $\gamma$, including both the multiphoton (for $\gamma \gg 1$) and tunneling (for $\gamma \ll 1$) regimes of ionization. According to such VG-SFA consideration, the calculated photoelectron momentum distributions (PMD) demonstrate a clear spatial asymmetry (viz., the photoelectron emission proved to be different along the opposite directions with respect to incident laser field polarization) even for particular bichromatic field(s) with polar symmetry of its total electric field strength. Moreover, a clear correlation between the spatial symmetry/asymmetry of photoelectron emission and that of applied vector potential of bichromatic field is reliably established, so that the symmetry of calculated PMD proved to follow precisely the symmetry of applied vector potential of incident laser bichromatic field, in a close analogy with the well-known Aharonov-Bohm effect. Based on these our findings, the attempt is undertaken to identify the physical mechanism underlying this intriguing phenomenon and provide it with a kind of semiclassical interpretation in terms of the partial contributions from photoelectron momenta in final continuum states to partial strong-field ATI amplidude(s) corresponding to a certain (fixed) number of either of two harmonics absorbed.

1. Introduction

A great interest to atomic and molecular photoprocesses – particularly, to such strong-field multiphoton processes as above threshold ionization (ATI) and/or high-order harmonic generation (HHG) – occurring in atomic and/or molecular species irradiated by a two-color (bichromatic) laser electromagnetic (EM) field consisting of coherent superposition of two commensurate harmonics of the same fundamental laser frequency does not seem to decrease for almost last three decades yet [1-17]. This is related to a number of the bichromatic field additional parameters (such as its harmonic frequency and intensity ratios, the relative phase shift between the harmonics etc.), which can potentially affect the dynamics and alter the direction of the photoprocesses (due to substantial changing its partial and total amplitude(s) and/or probability), thereby allowing for efficient coherent control of respective observables, such as e.g. the produced photoelectron angular and/or energy distributions (spectra) [18]. In other words, the bichromatic laser field can provide in principle for a more flexible and efficient tool (as compared to the monochromatic laser field) to affect and control a variety of laser-induced photoprocesses. An important

_______________________________________________________________

♣Corresponding author (e-mail address:  pavelpyakuz@gmail.com )

distinctive feature of the bichromatic fields is the presence of a spatial (polar) symmetry/asymmetry (in respect of opposite spatial directions) of the time dependence for its electric field strength $\mathbf{E}(t)$ and/or respective vector potential $\mathbf{A}(t)$. So, for a linearly polarized two-color EM field consisting of coherent superposition of the fundamental laser frequency $\omega$ and its second harmonic and relative phase shift $\varphi=0$ between these two colors there is a clear polar asymmetry for its electric field strength $\mathbf{E}(t)$ (i.e. spatial asymmetry with respect to the replacement $\mathbf{e} \rightarrow -\mathbf{e}$ of polarization vector $\mathbf{e}$) [1]. The influence of such a spatial symmetry/asymmetry of bichromatic fields on the produced photoelectron and photon emission has been discussed in numerous papers. In particular, the recent theoretical and experimental works [15-17] show that the photoelectron distributions produced in the atomic ATI in two-color circularly polarized laser fields can possess rotational or reflection symmetries depending on the symmetric properties of the laser electric fields and vector potentials. Similar studies with linearly polarized bichromatic fields [1-6] found that the photoelectron distributions are either polar symmetrical or asymmetrical depending on some field parameters, such as harmonic frequency and intensity ratios, the relative phase shifts etc. Moreover, it was revealed an interesting effect according to which the polar asymmetry of a photoelectron distribution correlates with the polar asymmetry of the vector-potential rather than electric field, pointing out that the vector-potential asymmetry has some physical meaning, by analogy to the well-known Aharonov-Bohm effect [19]. The physical mechanism underlying the polar asymmetry of the photoelectron distributions was explained in the frame of perturbation theory for EM interaction [6], but the nature of the above-mentioned correlation effect between the symmetry/asymmetry properties of the photoelectron distributions and vector-potentials has so far remained unclear. All this, together with the above-noted practical interest in the coherent control of different photoprocesses through changing and adjustment of the bichromatic field parameters makes the consideration of the problem is currently very relevant.

In the present paper, we consider the same problem of the polar asymmetry of the photoelectron emission due to strong-field ATI process in atoms irradiated by a linearly polarized bichromatic laser field consisting of coherent superposition of two commensurate harmonics of the same fundamental laser frequency $\omega$ under condition of arbitrary value of the so-called Keldysh parameter $\gamma$ and, particularly, for the least studied "intermediate" (between the multiphoton and tunneling) regimes of ionization. For this purpose, the problem is addressed within the conventional strong-field approximation (SFA) in standard velocity gauge (VG) formulation and the longitudinal photoelectron momentum distributions (PMD) are calculated for particular harmonic frequency ratios 1:2 ($\omega, 2\omega$) and 1:3 ($\omega, 3\omega$) at different relative phase shifts $\varphi$ between these harmonics. Based on such VG-SFA consideration, a clear correlation effect between the spatial symmetry/asymmetry properties of the applied vector potential $\mathbf{A}(t)$ and respective produced PMDs is established and proved as well as the attempt is undertaken to identify the physical mechanism underlying this intriguing phenomenon and provide it with a kind of semiclassical interpretation using the detailed analysis of respective partial two-color ATI strong-field amplitude(s) in terms of separate contributions from possible photoelectron momenta and/or energies acquired due to a certain (fixed) number of either of two harmonics absorbed in final continuum states.

## 2. Background theory of applied VG-SFA consideration

In this paper we consider the direct ATI of atoms in a linearly polarized bichromatic laser field with harmonic frequencies $\omega_1$ and $\omega_2$ which are multiples of the same fundamental laser frequency $\omega$ (i.e., $\omega_1 = s\omega$ and $\omega_2 = r\omega$, where $s$, $r$ are integers, and $s > r$)

$$\mathbf{E}(t) = \mathbf{e} E(t) = \mathbf{e}\left[ E_1 \sin(\omega_1 t) + E_2 \sin(\omega_2 t + \varphi) \right], \quad (1)$$

Here $\mathbf{e}$ is a unit polarization vector of the electric field $\mathbf{E}(t)$, $E_1$ and $E_2$ are the amplitudes of electric field strengths corresponding to either of colors, and $\varphi$ is a relative phase shift between the colors.

The respective vector potential $\mathbf{A}(t)$ associated with the two-color EM field (1) is

$$\mathbf{A}(t) = \mathbf{e}\, A(t) = \mathbf{e}\left[A_1 \cos(\omega_1 t) + A_2 \cos(\omega_2 t + \varphi)\right], \qquad (2)$$

where $A_1 = c E_1/\omega_1$ and $A_2 = c E_2/\omega_2$ are amplitudes of the harmonic vector potentials (here $c \approx 137$ is the light velocity in vacuum and, in addition, the atomic system of units $e = m = \hbar = 1$ is used throughout this presentation unless specially stated otherwise).

The Hamiltonian of EM interaction of the electron with incident laser EM field in the velocity gauge has the following form:

$$\hat{W}(\mathbf{r},t) = \frac{1}{c}\mathbf{A}(t)\,\hat{\mathbf{p}} + \frac{1}{2c^2}\mathbf{A}^2(t) \qquad (3)$$

According to the S-matrix formalism of standard SFA theory [20], the amplitude of the direct ATI process can be expressed through the S-matrix element of EM interaction $\hat{W}(\mathbf{r},t)$

$$S_{\mathbf{p}i} = -i\int_{-\infty}^{\infty} dt\, \langle \Psi_{\mathbf{p}}(\mathbf{r},t) | \hat{W}(\mathbf{r},t) | \Psi_i(\mathbf{r},t) \rangle, \qquad (4)$$

which corresponds to the electron transition from the initial bound state $\Psi_i(\mathbf{r},t) = \psi_i(\mathbf{r})\exp(iI_0 t)$ unperturbed by the laser field (with the binding energy $I_0$) to the final continuum state $\Psi_{\mathbf{p}}(\mathbf{r},t)$ with a canonical momentum $\mathbf{p}$ in the presence of the laser field only. The latter continuum wave function $\Psi_{\mathbf{p}}(\mathbf{r},t)$ is known as the Volkov's wave function [20, 21], which in the velocity gauge takes the form:

$$\Psi_{\mathbf{p}}(\mathbf{r},t) = (2\pi)^{-3/2} \cdot \exp\left[i\cdot\mathbf{p}\cdot\mathbf{r} - \frac{i}{2}\int_{-\infty}^{t} dt'\,(\mathbf{p}+\mathbf{k}_{t'})^2\right], \qquad (5)$$

where $\mathbf{k}_t = \int^t \mathbf{E}(t')\,dt' = \mathbf{A}(t)/c$ is a classical electron momentum due to the field [23]. In the case of the bichromatic field (1) the wave function (5) takes the form

$$\Psi_{\mathbf{p}}(\mathbf{r},t) = |\mathbf{p}\rangle \cdot \exp\left[-i\{(E_{\mathbf{p}} + U_p)t + \mathbf{p}\boldsymbol{\alpha}(t) + \beta(t) + \gamma(t)\}\right], \qquad (6)$$

where $E_{\mathbf{p}} = p^2/2$ is a photoelectron kinetic energy, $U_p = U_{p_1} + U_{p_2}$ is the photoelectron ponderomotive potential in the bichromatic laser field, $U_{p_j} = E_j^2/4\omega_j^2$, $j=1,2$ and $\boldsymbol{\alpha}(t)$, $\beta(t)$ and $\gamma(t)$ are periodic functions of time (for multiple harmonics $\omega_1$ and $\omega_2$)

$$\boldsymbol{\alpha}(t) = 2\mathbf{e}\left\{\sqrt{U_{p_1}}/\omega_1 \sin[\omega_1 t] + \sqrt{U_{p_2}}/\omega_2 \sin[\omega_2 t + \varphi]\right\}$$
$$\beta(t) = \left\{U_{p_1}/\omega_1 \sin[2\omega_1 t] + U_{p_2}/\omega_2 \sin[2(\omega_2 t + \varphi)]\right\}/2 \qquad (7)$$
$$\gamma(t) = 2\sqrt{U_{p_1} U_{p_2}}\left\{\sin[(\omega_2 - \omega_1)t + \varphi]/(\omega_2 - \omega_1) + \sin[(\omega_1 + \omega_2)t + \varphi]/(\omega_1 + \omega_2)\right\}$$

The time-dependent initial bound state $\Psi_i(\mathbf{r},t) = \psi_i(\mathbf{r})\exp(iI_0 t)$ is an initial laser-free wavefunction (orbital) for outermost atomic valence shell which is numerically composed using the respective Density Functional Theory (DFT) method applied along with a standard Gaussian basis set from the routines of GAUSSIAN software package [24].

Substituting the Volkov's wave function (6) into the amplitude (4) and performing some analytical transformations with Hermitian time-dependent operator of EM interaction (3), one obtains

$$S_{\mathbf{p}i} = -i\Phi_i(\mathbf{p})(E_{\mathbf{p}} + I_0)\int_{-\infty}^{\infty} dt\, B(\mathbf{p},t)\exp\left[i(E_{\mathbf{p}} + U_p + I_0)t\right], \qquad (8)$$

Here the following notations are introduced

$$\Phi_i(\mathbf{p}) = \langle \mathbf{p} | \psi_i(\mathbf{r}) \rangle \qquad (9)$$

for the spatial Fourier transform of the coordinate part $\psi_i(\mathbf{r})$ off initial unperturbed (laser-free) atomic wave-function, and

$$B(\mathbf{p},t) = \exp\left[-i\left(\mathbf{p}\boldsymbol{\alpha}(t) + \beta(t) + \gamma(t)\right)\right]. \tag{10}$$

For particular case of commensurate harmonics $\omega_1$ and $\omega_2$, the function $B(\mathbf{p},t)$ is a periodic function of time with a period $T = 2\pi/\Omega$ determined by the greatest common factor $\Omega$ of the harmonic frequencies $\omega_1$ and $\omega_2$. For example, $\Omega = \omega$ for the case of $\omega_1 = \omega$, $\omega_2 = 2\omega$, and $\Omega = 2\omega$ for the case of $\omega_1 = 2\omega$, $\omega_2 = 4\omega$. As a consequence the time-dependent function $B(\mathbf{p},t)$ can be expanded into the Fourier series of harmonics of the frequency $\Omega$:

$$B(\mathbf{p},t) = \sum_{n=-\infty}^{\infty} B(\mathbf{p},n)\exp(-in\Omega t), \quad B(\mathbf{p},n) = \frac{\Omega}{2\pi}\int_0^T dt\, B(\mathbf{p},t)\exp(in\Omega t) \tag{11}$$

where the time integration in (11) can be performed using standard methods for direct numerical integration. Substituting (11) into the amplitude (8) and performing analytical integration over time, one can derive

$$S_{\mathbf{p}i} = -2\pi i \sum_n F_i(\mathbf{p},n)\delta\left(E_{\mathbf{p}} + U_p + I_0 - n\Omega\right), \tag{12}$$

where $F_i(\mathbf{p},n)$ is the partial ionization amplitude corresponding to a certain (fixed) number $n$ of photons of the frequency $\Omega$ absorbed

$$F_i(\mathbf{p},n) = \left(E_{\mathbf{p}} + I_0\right)\Phi_i(\mathbf{p})B(\mathbf{p},n) \tag{13}$$

The differential ionization rate of the initial state $\Psi_i(\mathbf{r},t)$ can also be expressed through a sum over partial $n$-photon processes

$$dw^{(i)} = 2\pi \sum_{n=n_0}^{\infty} |F_i(\mathbf{p},n)|^2 \delta\left(E_{\mathbf{p}} + I_0 + U_p - n\Omega\right)\frac{d^3p}{(2\pi)^3}, \tag{14}$$

Here $n_0$ is the minimum number of absorbed photons of fundamental laser frequency $\omega$ necessary to overcome the binding potential, i.e. $n_0 = [(I_0 + U_p)/\omega] + 1$. The presence of the Dirac $\delta$-delta function (expressing the energy conservation in the process under consideration) dictates that each partial ($n$th) amplitude $F_i(\mathbf{p},n)$ in the sum of Eq. (14) is to be calculated only for respective definite value of the photoelectron final energy $E_{\mathbf{p}} = \mathbf{p}^2/2 = n\Omega - I_0 - U_p$. The latter means that the contributions only from open ATI channels (corresponding to $n \geq n_0$) are supposed to be taken into account under the summation over $n$ in (14).

The longitudinal photoelectron momentum distribution (PMD) due to ionization of the atomic valence shell $\psi_i(\mathbf{r})$ and corresponding to photoelectron emission with the longitudinal (along to the laser field polarization OZ axis) momentum component value within the interval between $p_\parallel$ and $p_\parallel + dp_\parallel$ can be found by means of integrating the respective differential ionization rate (14) over azimuthal angle $\varphi_p$ and magnitude $p$ of the photoelectron final momentum $\mathbf{p}$. Taking into account that $d^3p = p\,dp_\parallel\,d\phi_p$ in (14), one can obtain

$$\frac{dw^{(i)}(p_\parallel)}{dp_\parallel} = \frac{1}{(2\pi)^2}\sum_{n=n_0}^{\infty}\int_0^{2\pi}|F_i(p_\parallel,n)|^2 d\varphi_p, \tag{15}$$

Here the obvious equalities $\cos(\theta_p) = p_{\|}/p$, $\sin(\theta_p) = \sqrt{p^2 - p_{\|}^2}/p$ were used. Owing to the presence of the energy delta-function in (14) the integration over variable $p$ in (15) is just reduced to the replacement of $p$ by its value corresponding to a certain fixed number $n$ of photons absorbed

$$p(n) = \sqrt{2(n\Omega - I_0 - U_p)} \tag{16}$$

Finally, for atoms with a closed outermost valence shells, the momentum distributions (15) also implies a summation over all possible values of magnetic quantum number $m$

$$\frac{dw(p_{\|})}{dp_{\|}} = \sum_{m=-l}^{l} \frac{dw^{(m)}(p_{\|})}{dp_{\|}} \tag{17}$$

### 3. Main results and their discussion

In this section, we analyze an influence of the polar asymmetry of a linearly polarized bichromatic laser field (1) on the corresponding asymmetry of the electronemission produced in atomic ATI in such a field. For this purpose, the electron longitudinal momentum distributions for Ar atoms are calculated at harmonic frequencies $(\omega, 2\omega)$ and $(\omega, 3\omega)$ at different relative phase shifts $\varphi$. To simplify the calculations, the particular case of equal electric field amplitudes $E = E_1 = E_2$ ($A = A_1 = A_2$) of the harmonics is considered, which does not hide the essence of the problem. All calculations are performed within the conventional VG- SFA presented above, in the intermediate ionization regime characterized by the Keldysh parameter $\gamma \approx 1$, where $\gamma = \Omega\sqrt{2I_0}/E$ [25].

The presence of the spatial (polar) asymmetry of the linearly polarized bichromatic laser field (1) (i.e. the asymmetry with respect to the replacement $\mathbf{e} \to -\mathbf{e}$) is related to the different magnitudes of the local extrema (maxima) reached by the time-dependent function $E(t)$ along the opposite directions of the polarization plane (along the vector $\mathbf{e}$). The extent of such a polar asymmetry can be characterized by the asymmetry parameter [1]:

$$\xi_E^{(g)}(\varphi) = \left\langle \left(E(t)\right)^g \right\rangle, \tag{18}$$

where $g = s + r$ is the $g$ th order moment of the corresponding bichromatic EM field (1) and it is a function of the relative phase shift $\varphi$. Thus, the polar asymmetry of the time-dependent electric field strength $\mathbf{E}(t)$ is observed only for odd numbers of $g$ (i.e., when $s$ is odd, and $r$ is even, or vice versa), whereas, for even numbers of $g$ the electric field strength $\mathbf{E}(t)$ is always spatially symmetrical. Therefore, the asymmetry parameter (18) will be further used to evaluate if the total electric field strength $\mathbf{E}(t)$ of applied bichromatic EM field is spatially (polar) asymmetric or not. Analogously, the spatial (polar) asymmetry of the time-dependent vector potential $\mathbf{A}(t)$ (2) can be also evaluated by the respective asymmetry parameter

$$\xi_A^{(g)}(\varphi) = \left\langle \left(A(t)\right)^g \right\rangle, \tag{19}$$

The sign of the asymmetry parameters (18) and (19) determines in which half-plane within the polarization plane (negative or positive) the time-dependent function $E(t)$ and $A(t)$ reach the largest values. Let us denote the values of local maxima of the functions $E(t)$ and $A(t)$ reached in these half-planes as $E^{(+)}$, $A^{(+)}$ and $E^{(-)}$, $A^{(-)}$, respectively. If the parameters (18) and (19) are zero, the bichromatic field is completely symmetric ($E^{(+)} = E^{(-)}$ and $A^{(+)} = A^{(-)}$). Whereas, if one of the parameters (18) and (19) is different from zero, the corresponding vectors $\mathbf{E}(t)$ and $\mathbf{A}(t)$ have the polar asymmetry due to inequality of the maxima, viz., if $\xi_{E,A}^{(g)}(\varphi)<0$ then $E^{(+)}<E^{(-)}$ ($A^{(+)}<A^{(-)}$), while if $\xi_{E,A}^{(g)}(\varphi)>0$ then $E^{(+)}>E^{(-)}$ ($A^{(+)}>A^{(-)}$).

In Figure 1, the polar asymmetry of the vectors $\mathbf{E}(t)$ and $\mathbf{A}(t)$ is illustrated for the case of bichromatic field consisting of odd and even harmonics $(\omega,2\omega)$ of a linear polarization. The asymmetry parameters (18) and (19) for that field are equal $\xi_E^{(3)}(\varphi)=-3E^3\sin(\varphi)/4$ and $\xi_A^{(3)}(\varphi)=3A^3\cos(\varphi)/4$. Figure 1(a) shows that at $\varphi=0$, the electric-field vector $\mathbf{E}(t)$ is symmetric ($E^{(+)}=E^{(-)}$, as $\xi_E^{(3)}(0)=0$), while the corresponding vector potential $\mathbf{A}(t)$ is asymmetric ($A^{(+)}>A^{(-)}$, as $\xi_A^{(3)}(0)>0$). At $\varphi=\pi/2$, the picture becomes the opposite to the case of $\varphi=0$ (Figure 1(b)). In the case of $\varphi=\pi$, the electric field $\mathbf{E}(t)$ becomes symmetric again ($\xi_E^{(3)}(\pi)=0$), while the polar asymmetry of the vector potential $\mathbf{A}(t)$ is reversed ($A^{(+)}<A^{(-)}$, as $\xi_A^{(3)}(\pi)<0$). The bichromatic fields consisting of only odd harmonics, such as $(\omega,3\omega)$, are always symmetrical at any phase shifts $\varphi$ and, therefore, there is no need to present them in a separate figure.

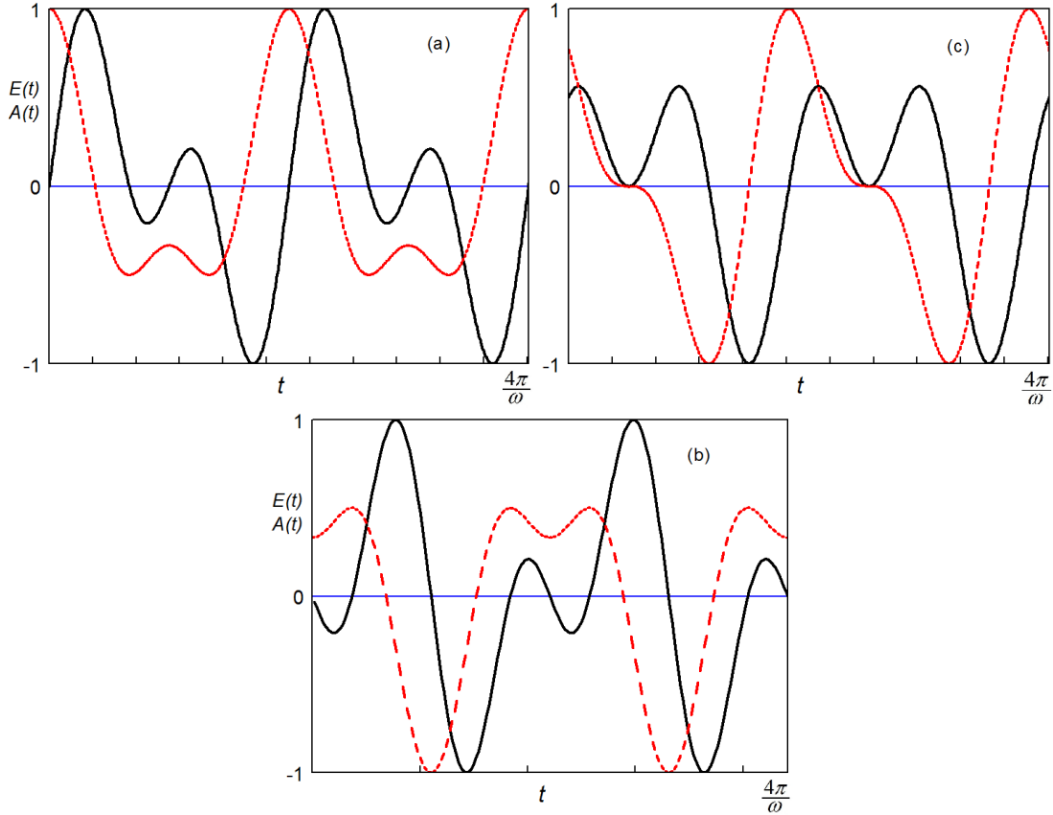

**Figure 1.** The time dependence of electric field strength $E(t)$ (solid lines), and the corresponding vector potential $A(t)$ [dashed (red) lines] of a linearly polarized bichromatic EM field with harmonic frequencies $(\omega,2\omega)$ and equal harmonic electric field strengths $E=E_1=E_2$ at relative phase shifts (*a*) $\varphi=0$, (*b*) $\varphi=\pi/2$ and (*c*) $\varphi=\pi$. For the convenience the presented results are normalized to unit. (This figure is in color only in the electronic version)

Let us now consider the calculated longitudinal momentum distributions for photoelectron emission from Ar atoms having the outermost valence shell of or $p$-type. The respective ionization rate (17) can be reduced to a sum of three terms corresponding to separate contributions from three outermost atomic valence orbitals with respective possible values of magnetic quantum number $m = -1, 0, 1$:

$$\frac{dw(p_{\|})}{dp_{\|}} = \frac{dw^{(m=-1)}(p_{\|})}{dp_{\|}} + \frac{dw^{(m=0)}(p_{\|})}{dp_{\|}} + \frac{dw^{(m=1)}(p_{\|})}{dp_{\|}} \quad (20)$$

The polar asymmetry of the distribution (20) can be also analyzed and identified according to the unequal values of its peak (maximum) magnitude $dw_{\max}(+p_{\|})$ and $dw_{\max}(-p_{\|})$ reached in the positive and negative directions within the laser field polarization plane, respectively. In particular, if $dw_{\max}(+p_{\|}) = dw_{\max}(-p_{\|})$, the momentum distributions are acknowledged as symmetrical. If $dw_{\max}(+p_{\|}) > dw_{\max}(-p_{\|})$, they are shifted to the positive direction (along the OZ axis within the laser field polarization plane), while if $dw_{\max}(+p_{\|}) < dw_{\max}(-p_{\|})$ - to the negative direction of the OZ axis. Figure 2 illustrates the longitudinal PMDs calculated for Ar atoms ionized by a bichromatic laser field consisting of harmonic frequencies $(\omega, 2\omega)$ at different relative phase shifts $\varphi$. The symmetry/asymmetry of the field $(\omega, 2\omega)$ has been analyzed above (see Figure 1). According to a general physical considerations, the momentum distribution at $\varphi = 0$ is expected to be symmetrical in the positive and negative directions within the laser field polarization plane, Contrary to such quite a natural prediction, Figure 2 demonstrates that the PMD calculated for $\varphi = 0$ is asymmetric and shifted to the negative direction of the OZ axis ($dw_{\max}(-p_{\|}) > dw_{\max}(+p_{\|})$). Conversely, for relative phase $\varphi = \pi/2$, the calculated PMD

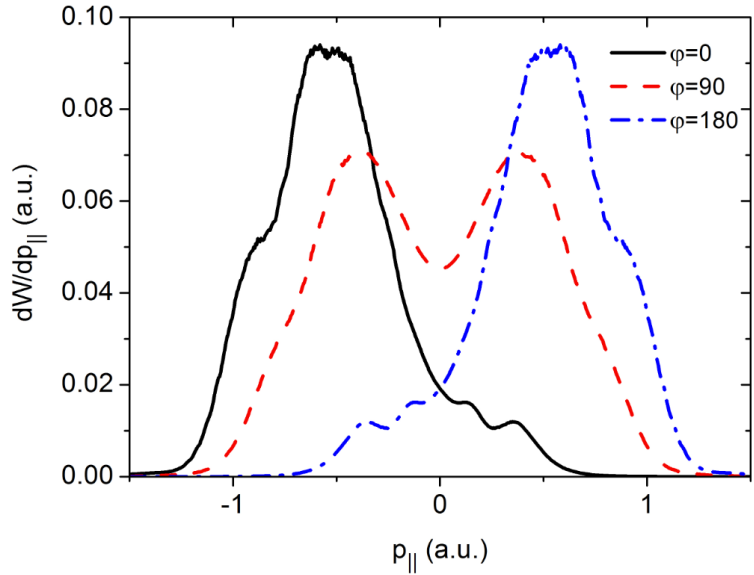

**Figure 2.** The longitudinal momentum distributions of electrons emitted from Aratoms in ATI produced by linearly polarized bichromatic EM field with harmonic frequencies $(\omega, 2\omega)$ of the fundamental wavelength $\lambda = 800$ nm and equal harmonic intensities $I_1 = I_2 = 10^{14}$ W/cm$^2$ for which the value of the Keldysh parameter $\gamma \approx 1.15$. (This figure is in color only in the electronic version)

becomes clearly symmetrical $(dw_{\max}(+p_\parallel) = dw_{\max}(-p_\parallel))$, in contrast to the asymmetry of the respective electric field strength $\mathbf{E}(t)$ ($\xi_E^{(3)}(\pi/2) < 0$, $E^{(+)} < E^{(-)}$) of applied bichromatic laser field.

For both cases of $\varphi = \pi$ and $\varphi = 0$, the calculated PMDs are asymmetrical despite the obviously seen polar symmetry of the respective electric field strength $\mathbf{E}(t)$ ($E^{(+)} = E^{(-)}$, $\xi_E^{(3)}(\pi) = 0$), although with the only difference that at $\varphi = \pi$ the respective PMD is shifted to the positive direction of the OZ axis $(dw_{\max}(-p_z) < dw_{\max}(+p_z))$. On the other hand, when summarizing these three cases, one can also note a clear and remarkable correlation between the polar asymmetry of the calculated PMDs and the polar asymmetry of the respective vector potential of applied bichromatic laser field, viz. they are both either symmetric or asymmetric for the same values of $\varphi$. More specifically, both the distribution $dw(p_\parallel)/dp_\parallel$ and $\mathbf{A}(t)$ are asymmetrical at $\varphi = 0$ (viz., $dw_{\max}(-p_\parallel) > dw_{\max}(+p_\parallel)$ and $A^{(-)} < A^{(+)}$); whereas, they are both symmetrical at $\varphi = \pi/2$ (viz., $dw_{\max}(-p_\parallel) = dw_{\max}(+p_\parallel)$ and $A^{(-)} = A^{(+)}$); while they become again asymmetrical ($dw_{\max}(-p_z) < dw_{\max}(+p_z)$ and $A^{(-)} > A^{(+)}$) at $\varphi = \pi$. Thus, the photoelectron emission proved to be sensitive to the spatial symmetry/asymmetry of the applied EM field vector potential $\mathbf{A}(t)$, rather than that of the respective associated electric field strength $\mathbf{E}(t)$ (which might be expected to be only physically meaningful). The latter phenomenon seems to be quite a surprising and intriguing that allowed the authors of [4] to suppose that the phenomenon under consideration is a remarkable (though, a very rare) pattern, for which the properties of the vector potential $\mathbf{A}(t)$ have a superiority over those of the associated (and always only meaningful for various different physical effects) electric field strength $\mathbf{E}(t)$, in a close analogy with the well-known Aharonov-Bohm effect [19].

The physical mechanism underlying the polar asymmetry of the electron momentum distributions in linearly polarized bichromatic laser fields was given in [5] in the framework of the perturbation theory for EM interaction. Namely, such phenomenon was explained in terms of the quantum interference between all contributing transition amplitudes available in the bichromatic fields. For example, in the bichromatic fields consisting of odd and even harmonics, such as $(\omega, 2\omega)$, the electron can make a transition to the final state with the same energy, but different parities. A quantum interference between such EM transitions results in the spatial asymmetry of the PMDs even for the bichromatic fields with symmetric electric field strength $\mathbf{E}(t)$. Meantime, the similar interpretation of the phenomenon in terms of the quantum interference seems to be very complicated and non-transparent within the framework of the general S-matrix formalism of SFA approach currently applied in this paper, so that an additional (though ) attempt to provide the phenomenon with a different clear semiclassical interpretation is made here.

In particular, to understand the reason of the polar asymmetry of the electron emission in a spatially symmetrical bichromatic field(s), we suggest to analyze a time dependence of the classical momentum $\mathbf{k}_t = \int^t \mathbf{E}(t')dt' = \mathbf{A}(t)/c$ acquired by photoelectron due to EM interaction with the bichromatic electric field $\mathbf{E}(t)$ and estimate its contribution to the total electron energy $(\mathbf{p} + \mathbf{k}_t)^2/2$ during the laser field period $T$. In the case of the bichromatic field with frequencies $(\omega, 2\omega)$, the momentum is $\mathbf{k}_t = \mathbf{k}_\omega(t) + \mathbf{k}_{2\omega}(t)$, here $\mathbf{k}_\omega(t)$ and $\mathbf{k}_{2\omega}(t)$ are momenta acquired by electron due to its interaction with the EM fields of the first and second harmonics, respectively. The latter interactions are determined by the harmonic electric fields

$\mathbf{E}_1(t) = \mathbf{e} E_1 \sin(\omega t)$ and $\mathbf{E}_2(t) = \mathbf{e} E_2 \sin(2\omega t)$, and, in addition, proportional to associated vector potentials $\mathbf{A}_1(t)$ and $\mathbf{A}_2(t)$:

$$\mathbf{k}_\omega(t) = \int^t \mathbf{E}_1(t) dt' = \frac{\mathbf{A}_1(t)}{c} = \mathbf{e} \frac{E_1}{\omega} \cos(\omega t)$$

$$\mathbf{k}_{2\omega}(t) = \int^t \mathbf{E}_2(t) dt' = \frac{\mathbf{A}_2(t)}{c} = \mathbf{e} \frac{E_1}{2\omega} \cos(2\omega t)$$

One can see that even for equal harmonic electric field amplitudes $E_1 = E_2 = E$, the magnitudes of the momenta $\mathbf{k}_\omega(t)$ and $\mathbf{k}_{2\omega}(t)$ are not equal ($|\mathbf{k}_\omega(t)| = 2|\mathbf{k}_{2\omega}(t)|$).

Figure 3 illustrates the time dependence of the momentum projections $\mathbf{k}_t$, $\mathbf{k}_\omega(t)$ and $\mathbf{k}_{2\omega}(t)$ on the OZ axis (along the polarization vector $\mathbf{e}$) at $\varphi = 0$. One can see that during the fundamental laser field period $T = 2\pi/\omega$ the momenta $\mathbf{k}_\omega(t)$ and $\mathbf{k}_{2\omega}(t)$ can be co- and counter-directional to each other, resulting in symmetry/asymmetry (with equal/unequal projections) of the total kinetic momentum $\mathbf{k}_t$ along the opposite directions within the laser field polarization plane. Namely, the projection of $\mathbf{k}_t$ has the greatest value on the positive direction at $t = 0$ and $t = 2\pi/\omega$, as $\mathbf{k}_\omega(t)$ and $\mathbf{k}_{2\omega}(t)$ coincide in direction and have their maximum values only at these moments of time. The latter, in its turn, means that the total photoelectron energy $(\mathbf{p} + \mathbf{k}_t)^2/2$ gained along the positive direction will be larger as compared to that along the negative direction, even in bichromatic fields with symmetrical electric field strength $\mathbf{E}(t)$ (see Figure 1(a)).

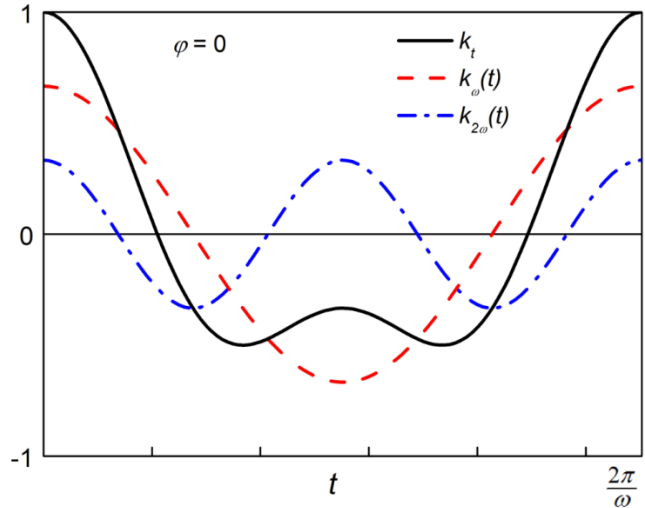

**Figure 3.** The time dependence of projections of the momenta $\mathbf{k}_t$, $\mathbf{k}_\omega$ and $\mathbf{k}_{2\omega}$ on the OZ axis (along the polarization vector $\mathbf{e}$) in the linearly polarized bichromatic EM field with parameters as in figure 1 (a). The results are normalized to unit. (This figure is in color only in the electronic version)

Thus, one can conclude that when an electron absorbs a certain number of photons and, therefore, has a certain final energy, it is energetically more profitable for it to be emitted to the negative direction of the OZ axis, than in the positive, which is reflected in the polar asymmetry of the corresponding momentum distribution at $\varphi = 0$ (see Figure 2). Taking into account also that the momenta $\mathbf{k}_t$, $\mathbf{k}_\omega(t)$ and $\mathbf{k}_{2\omega}(t)$ are proportional to the vector potentials $\mathbf{A}(t)$, $\mathbf{A}_1(t)$ and $\mathbf{A}_2(t)$, the physical mechanism underlying the above-mentioned paradoxical correlation between the polar asymmetry of the vector potential $\mathbf{A}(t)$ and the produced photoelectron momentum distributions becomes understandable.

Figure 4 illustrates the time dependence of the projections $\mathbf{k}_t$, $\mathbf{k}_\omega(t)$ and $\mathbf{k}_{2\omega}(t)$ on the axis OZ at $\varphi = \pi/2$. It can be seen that during the field period $T = 2\pi/\omega$ the momentum $\mathbf{k}_t$ is symmetrical in relation to the equality of its maximum projections on the opposite directions. As a result, one can therefore expect that the total photoelectron energy $(\mathbf{p}+\mathbf{k}_t)^2/2$ gained along the opposite spatial directions are to be equal even in bichromatic field(s) with asymmetric electric field $\mathbf{E}(t)$ (see Figure 1(b)). In other words, for a certain fixed number of absorbed photons, the two opposite spatial directions of photoelectron emission are to be energetically equivalent, so that it can be emitted along opposite directions with equal probabilities resulting in the polar symmetry of the corresponding PMD at $\varphi = \pi/2$ (see Figure 2). The case of $\varphi = \pi$ can be analyzed in the similar way.

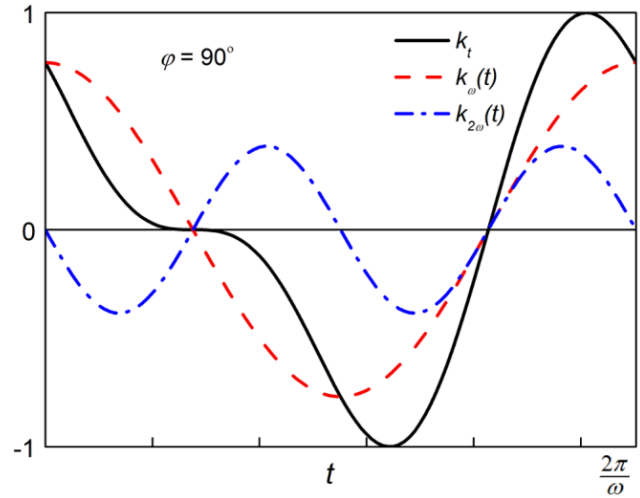

**Figure 4.** The time dependence of projections of the momenta $\mathbf{k}_t$, $\mathbf{k}_\omega$ and $\mathbf{k}_{2\omega}$ on the OZ axis (along the polarization vector $\mathbf{e}$) in the linearly polarized bichromatic EM field with parameters as in figure 1 (b). The results are normalized to unit.

The same analysis of partial contributions to calculated PMDs from the photoelectron momenta acquired in final continuum states due to EM interaction with either of two-color harmonics can be also made for bichromatic field(s) consisting of odd harmonics only, such as $(\omega, 3\omega)$, and such analysis proved that the total photoelectron energy $(\mathbf{p}+\mathbf{k}_t)^2/2$ gained along the opposite directions within the laser field polarization plane are to be always equal to each other for any phase shifts $\varphi$, thereby resulting in the polar symmetry of the corresponding calculated PMDs (see Figure 5).

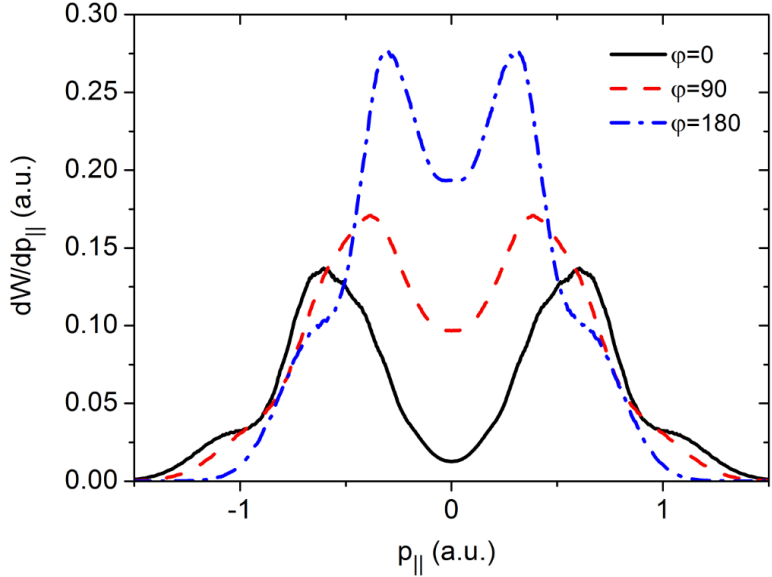

**Figure 5.** The longitudinal photoelectrons momentum distributions produced in ATI of Ar atoms irradiated by linearly polarized bichromatic EM field for harmonic frequencies $(\omega, 3\omega)$ of the fundamental wavelength $\lambda = 800$ nm and for equal harmonic intensities $I_1 = I_2 = 10^{14}$ W/cm$^2$. The value of the Keldysh parameter $\gamma \approx 1.15$. (This figure is in colour only in the electronic version)

## 4. Conclusions

We have theoretically studied the problem of the spatial (polar) asymmetry of photoelectron emission produced due to atomic ATI in a linearly polarized two-color laser field consisting of commensurate harmonics (of frequencies $\omega_1$ and $\omega_2$), which are multiples of the same fundamental laser frequency $\omega$. Within the fully quantum-mechanical consideration of S-matrix formalism of conventional VG-SFA approach, the longitudinal photoelectron momentum distributions (PMD) produced in the bichromatic fields consisting of harmonic frequencies $(\omega, 2\omega)$ and $(\omega, 3\omega)$ at different relative phase shifts $\varphi$ were calculated under conditions of arbitrary ionization regimes (including both multiphoton and tunneling ones). Our VG-SFA based calculation results confirm the conclusions of earlier works that for the fields with odd and even harmonics, such as $(\omega, 2\omega)$, the photoelectron emission can possess the polar asymmetry depending on the relative phase shift between the harmonics, whereas, for bichromatic field(s) consisting of odd harmonics only, such as $(\omega, 3\omega)$, the resulting PMDs are to be always symmetrical at any value of relative phase shift.

A significant role of partial contributions from the photoelectron momenta $\mathbf{k}_{s\omega}(t)$ and $\mathbf{k}_{r\omega}(t)$ acquired in final continuum states due to EM interaction with electric fields $\mathbf{E}_1(t)$ and $\mathbf{E}_2(t)$ of either harmonic has been revealed in the formation of produced PMD and its spatial symmetry. During the field period $T = 2\pi/\omega$ the momenta $\mathbf{k}_{s\omega}(t)$ and $\mathbf{k}_{r\omega}(t)$ can be either co- or counter-directional relative to each other, that results in symmetry/asymmetry (with equal/unequal values of projections) of the total kinetic momentum $\mathbf{k}_t = \mathbf{k}_{s\omega}(t) + \mathbf{k}_{r\omega}(t)$ along the opposite directions within the laser field polarization plane. As a result, the polar asymmetry of the photoelectron emission, in particular, is demonstrated to arise due to the fact that the final photoelectron energies gained along the opposite directions are different (not equal to each other) and, thus, only one of the opposite directions becomes more energetically favorable under absorption of a certain number of photons (energy) even in the case of bichromatic field(s) with polar symmetry of total electric field strength.

Taking into account also that the momentum $\mathbf{k}_t$ is proportional to the vector potential $\mathbf{A}(t)$, the physical mechanism underlying the above-mentioned paradoxical correlation between the spatial (polar) symmetry/asymmetry of produced PMD and that of respective vector potential $\mathbf{A}(t)$ of applied bichromatic field seems to become more transparent and understandable.

## Acknowledgements

The authors express their gratitude and acknowledgements to Dr, B.A.Abdullaev for helpful comments and valuable remarks under discussion of the results reported in the paper. The research described in this presentation was also made possible in part by financial support (Grant No. **Ф2-60**) from the State Committee of Science and Technologies of Republic of Uzbekistan.